\documentstyle[12pt]{article}
\input{epsf}
\catcode`\@=11

\@addtoreset{equation}{section}
\def\theequation{\arabic{section}.\arabic{equation}}
\catcode`\@=11
\def\section{\@startsection{section}{1}{\z@}{3.5ex plus 1ex minus
   .2ex}{2.3ex plus .2ex}{\large\bf}}

\def\eqnarray{\let\@currentlabel=\theequation\refstepcounter{equation}
    \global\@eqnswtrue
    \global\@eqcnt\z@\tabskip\@centering\let\\=\@eqncr
    $$\halign to \displaywidth\bgroup\@eqnsel\hskip\@centering
      $\displaystyle\tabskip\z@{##}$&\global\@eqcnt\@ne
       \hfil${{}##{}}$\hfil
      &\global\@eqcnt\tw@ $\displaystyle\tabskip\z@{##}$\hfil
       \tabskip\@centering&\llap{##}\tabskip\z@\cr}
\def\lefteqn#1{\hbox to 4\arraycolsep{$\displaystyle #1$\hss}}
\def\thesection{\arabic{section}.}

\def\appendix{\setcounter{section}{0}
        \def\thesection{Appendix.}
        \def\theequation{\Alph{section}.\arabic{equation}}}

\long\def\@makefntext#1{\parindent 0cm\noindent
\hbox to 1em{\hss$^{\@thefnmark}$}#1}

\def\IR{{\hbox{{\rm I}\kern-.2em\hbox{\rm R}}}}
\def\IH{{\hbox{{\rm I}\kern-.2em\hbox{\rm H}}}}
\def\IC{{\ \hbox{{\rm I}\kern-.6em\hbox{\bf C}}}}
\def\IZ{{\hbox{{\rm Z}\kern-.4em\hbox{\rm Z}}}}
\def\rref#1{(\ref{#1})}
\def\sp{\hskip -.08em\relax}

\newcommand{\beq}{\begin{equation}}
\newcommand{\eeq}{\end{equation}}

\newcommand{\NPB}[1]{{\sl Nucl.~Phys.} {\bf B#1}}
\newcommand{\Ann}[1]{{\sl Ann.~Phys.} {\bf #1}}
\newcommand{\CMP}[1]{{\sl Commun.\ Math.\ Phys.} {\bf #1}}
\newcommand{\PLB}[1]{{\sl Phys.~Lett.} {\bf B#1}}

\newcommand{\PRL}[1]{{\sl Phys.~Rev.\ Lett.} {\bf #1}}

\newcommand{\PRD}[1]{{\sl Phys.~Rev.} {\bf D#1}}

\begin{document}
%%%%%%%%%%%%%%%%%%%%%%%%%%%%%%%%%%%%%%%%%%%%%%%%%%%%%%%%%%%%%%%%%%%%%%%%%%%
%     C I T E . S T Y
%     compressed lists of numerical citations: [11-16]
%     see also OVERCITE.STY and DRFTCITE.STY
%
%     Copyright (C) 1989-1992 by Donald Arseneau
%     These macros may be freely transmitted, reproduced, or modified for
%     non-commercial purposes provided that this notice is left intact.
%
%
%  \@citen contains the code that parses the list of names, ignoring
%  spaces after commas, writes the aux file \citation, and formats the
%  number list.  \citen can be used by itself to give citation numbers
%  without the other formatting; e.g., "See also ref.~\citen{junk}."
%
\def\citen#1{%
\edef\@tempa{\@ignspaftercomma,#1, \@end, }% ignore spaces in parameter list
\edef\@tempa{\expandafter\@ignendcommas\@tempa\@end}%
\if@filesw \immediate \write \@auxout {\string \citation {\@tempa}}\fi
\@tempcntb\m@ne \let\@h@ld\relax \let\@citea\@empty
\@for \@citeb:=\@tempa\do {\@cmpresscites}%
\@h@ld}
%
% for ignoring spaces in the input:
\def\@ignspaftercomma#1, {\ifx\@end#1\@empty\else
   #1,\expandafter\@ignspaftercomma\fi}
\def\@ignendcommas,#1,\@end{#1}
%
% For each citation, check if it is defined, if it is a number, and
% if it is a consecutive number that can be represented like 3-7.
%
\def\@cmpresscites{%
 \expandafter\let \expandafter\@B@citeB \csname b@\@citeb \endcsname
 \ifx\@B@citeB\relax % undefined
    \@h@ld\@citea\@tempcntb\m@ne{\bf ?}%
    \@warning {Citation `\@citeb ' on page \thepage \space undefined}%
 \else%  defined
    \@tempcnta\@tempcntb \advance\@tempcnta\@ne
    \setbox\z@\hbox\bgroup % check if citation is a number:
    \ifnum\z@<0\@B@citeB \relax
       \egroup \@tempcntb\@B@citeB \relax
       \else \egroup \@tempcntb\m@ne \fi
    \ifnum\@tempcnta=\@tempcntb % Number follows previous--hold on to it
       \ifx\@h@ld\relax % first pair of successives
          \edef \@h@ld{\@citea\@B@citeB}%
       \else % compressible list of successives
%         % use \hbox to avoid easy \exhyphenpenalty breaks
          \edef\@h@ld{\hbox{--}\penalty\@highpenalty \@B@citeB}%
       \fi
    \else   %  non-successor--dump what's held and do this one
       \@h@ld \@citea \@B@citeB \let\@h@ld\relax
 \fi\fi%
 \let\@citea\@citepunct
}
%
%%    To put space after the comma, use:
\def\@citepunct{,\penalty\@highpenalty\hskip.13em plus.1em minus.1em}%
%%    For no space after comma, use:
%% \def\@citepunct{,\penalty\@highpenalty}%
%%
%
%  Make \@citex refer to \citen:
%
\def\@citex[#1]#2{\@cite{\citen{#2}}{#1}}%
%
%  Replacement for \@cite.  Give one normal space before the citation,
%  set high penalties for linebreaks,
%
\def\@cite#1#2{\leavevmode\unskip
  \ifnum\lastpenalty=\z@ \penalty\@highpenalty \fi % highpenalty before
  \ [{\multiply\@highpenalty 3 #1% % triple-highpenalties within list
      \if@tempswa,\penalty\@highpenalty\ #2\fi % and before note.
    }]\spacefactor\@m}
\let\nocitecount\relax  % in case \nocitecount was used for drftcite
%
%%%%%%%%%%%%%%%%%%%%%%%%%%%%%%%%%%%%%%%%%%%%%%%%%%%%%%%%%%%%%%%%%%%%%%%%%%
\begin{titlepage}
\vspace{.5in}
\begin{flushright}
UCD-95-30\\
gr-qc/9509024\\
September 1995\\
\end{flushright}
\vspace{.5in}
\begin{center}
{\Large\bf
 Statistical Mechanics\\[2ex] and Black Hole Entropy}\\
\vspace{.4in}
{S.~C{\sc arlip}\footnote{\it email: carlip@dirac.ucdavis.edu}\\
       {\small\it Department of Physics}\\
       {\small\it University of California}\\
       {\small\it Davis, CA 95616}\\{\small\it USA}}
\end{center}

\vspace{.5in}
\begin{center}
\begin{minipage}{4in}
\begin{center}
{\large\bf Abstract}
\end{center}
{\small
I review a new (and still tentative) approach to black hole
thermodynamics that seeks to explain black hole entropy
in terms of microscopic quantum gravitational boundary
states induced on the black hole horizon.  (Talk given at
CAM '95, joint meeting of the Canadian Association of Physicists,
the American Physical Society, and the Mexican Physical Society,
Quebec City, Canada.)
}
\end{minipage}
\end{center}
\end{titlepage}
\addtocounter{footnote}{-1}

It has been over twenty years since we first learned from Bekenstein
\cite{Bek} and Hawking \cite{Hawk} that black holes are thermodynamic
systems, characterized by temperatures and entropies.  But despite
considerable progress in the field, we still lack a convincing
``statistical mechanical'' picture of black hole thermodynamics.
Indeed, black hole entropy remains rather fundamentally paradoxical.
On the one hand, given the macroscopic parameters of mass, charge,
and angular momentum, a black hole configuration has the highest
obtainable entropy, implying at least naively that the black hole
has a large number of macroscopically indistinguishable microscopic
states.  On the other hand, a black hole has no hair: given the same
macroscopic parameters, there is, in fact, only one classical black
hole state.

A number of attempts have been made to resolve this paradox (see
\cite{Rev} for a review), but none is yet generally accepted.  In
this paper, I would like to advocate a new, still rather tentative
approach, which seeks to explain black hole entropy in terms of
microscopic quantum gravitational states on the horizon.  I do not
yet know how to apply this picture to realistic (3+1)-dimensional
black holes, but at least in the simpler case of 2+1 dimensions, it
has been shown (modulo some reasonable assumptions about quantization)
to lead to the correct entropy \cite{Carbh}.

My starting point is the observation that any quantum mechanical
statement about black holes is necessarily a statement about {\em
conditional\/} probabilities: for instance, ``If spacetime contains
an event horizon of a certain size, then we should see Hawking
radiation with a certain spectrum.''  So the first question we
must ask is how to impose such a condition---a restriction on the
geometry of a spacetime hypersurface---in a quantum mechanical
computation.  One obvious answer is to start with a path integral
formalism, split spacetime $M$ along a hypersurface $\Sigma$ into
two pieces, say $M_1$ and $M_2$, and perform separate path integrals
over $M_1$ and $M_2$ with suitable boundary conditions on $\Sigma$
(figure 1).  We are therefore naturally led to consider the problem
of ``sewing'' path integrals.

\begin{figure}[h]
\begin{center}
\leavevmode
\epsfxsize=3.7in
\epsfbox{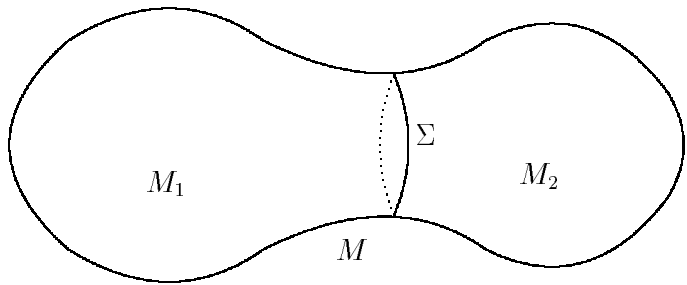}
\end{center}
\vspace{-4ex}
\caption{The manifold $M$ is formed by ``sewing'' $M_1$ and $M_2$
along $\Sigma$.}
\end{figure}
\vspace{1ex}

In the remainder of this paper, I will deal with three successively
more complicated models: a scalar field, an abelian Chern-Simons theory,
and (2+1)-dimensional gravity.  I will  conclude with a discussion
of possible generalization to realistic (3+1)-dimensional gravity
and implications for black hole thermodynamics.

\section{Sewing Scalar Fields}

Let us begin with a simple example, which nevertheless can teach us
some important lessons.  Consider a scalar field $\phi$ on $M$, with
an action
\beq
I_M[\phi] = {1\over2} \int_M \!d^n\sp x\sqrt{-g}\,\phi\Delta\phi
\label{a1}
\eeq
and a partition function
\beq
Z[M] = \int [d\phi]\, e^{iI_M[\phi]} = {\det}^{-1/2}\Delta_M .
\label{a2}
\eeq
Splitting $M$ along $\Sigma$, one might hope to find a relationship
of the form
\beq
Z[M] = Z[M_1]Z[M_2] ,
\label{a3}
\eeq
which would require that
\beq
\det\Delta_M = \det\Delta_{M_1}\det\Delta_{M_2} .
\label{a4}
\eeq
Strictly speaking, equation \rref{a4} doesn't quite make sense, since
determinants on $M_1$ and $M_2$ require boundary conditions for their
definition.  But in fact, \rref{a4} does not hold for any choice of
boundary conditions.

It is clear where we have gone wrong: the action \rref{a1} is simply
not correct for a manifold with boundary.  In fact, $I_{M_1}[\phi]$
has no classical extrema for generic boundary values: its variation is
\beq
\delta I_{M_1}[\phi] = \int_{M_1} \!d^n\sp x\sqrt{-g}\,\delta\phi\Delta\phi
   + {1\over2}\int_\Sigma \!d^{n-1}\sp x\sqrt{h}\, \left( \phi\,
   n^\mu\nabla_{\!\mu}\delta\phi-\delta\phi\,n^\mu\nabla_{\!\mu}\phi\right),
\label{a5}
\eeq
and the surface term does not vanish for either Dirichlet or Neumann
boundary conditions.  (Here $h$ is the induced metric on $\Sigma$, and
$n^\mu$ is the unit normal.)  The cure is obvious, however---if  we
choose boundary conditions in which $\phi$ is fixed at the boundary,
for instance, we must add to the action a surface term
\beq
I_\Sigma[\phi]= -{1\over2}\int_\Sigma \!d^{n-1}\sp x\sqrt{h}\,
   \phi\,n^\mu\nabla_{\!\mu}\phi
\label{a6}
\eeq
to obtain a total action
\beq
I_{M_1}'[\phi] = I_{M_1}[\phi] + I_\Sigma[\phi] = -{1\over2}\int_M
   \!d^n\sp x\sqrt{-g}\,\,\nabla^\mu\phi\,\nabla_{\!\mu}\phi .
\label{a7}
\eeq
(If we choose instead to fix the normal derivative of $\phi$ at $\Sigma$,
we should take as our action $I_{M_1}''[\phi] = I_{M_1}[\phi] -
I_\Sigma[\phi]$.)

The computation of $Z[M_1]$ is now standard.  If we denote the boundary
value of $\phi$ by $\phi_0$, and let $\bar\phi$ be the classical solution
with boundary value $\phi_0$, then
\beq
Z[M_1][\phi_0] = \int [d\phi] \exp\left\{iI_{M_1}'[\phi]\right\}
   = \exp\left\{iI_{M_1}'[\bar\phi]\right\}\,{\det}^{-1/2}\Delta_{M_1} ,
\label{a8}
\eeq
where the determinant is taken with Dirichlet boundary conditions.  Note
that the classical action $I_{M_1}'[\bar\phi]$ may be written as a bilinear
functional of $\phi_0$,
\beq
I_{M_1}'[\bar\phi] = \int_\Sigma \!d^{n-1}\sp x\sqrt{h(x)}
 \int_\Sigma \!d^{n-1}\sp x'\sqrt{h(x')}\,\,\phi_0(x)K_{M_1}(x,x')\phi_0(x')
 {}.
\label{a9}
\eeq
The kernel $K(x,x')$ is known as the Poisson kernel \cite{Forman}; it
may be written in terms of normal derivatives of the Greens function
for the Laplacian $\Delta_{M_1}$, and has a straightforward generalization
to other field theories.

We can now ``sew'' $M_1$ and $M_2$ by integrating over boundary values
$\phi_0$:
\beq
Z[M] = \int [d\phi_0] Z[M_1][\phi_0]Z[M_2][\phi_0] .
\label{a10}
\eeq
The integral over $\phi_0$ is again Gaussian, and can be performed
exactly; we find that \rref{a4} must be replaced by
\beq
\det\Delta_M = \det\Delta_{M_1}\det\Delta_{M_2}\det(K_{M_1}+K_{M_2}) .
\label{a11}
\eeq
This relationship among determinants, and its generalization to a wide
variety of free field theories, has been rigorously demonstrated to be
true \cite{CCDD}.

None of this is surprising, of course, but it does demonstrate two
crucial lessons:
\begin{enumerate}
\item An action appropriate for a closed manifold may not be suitable
for a manifold with boundary;
\item The precise form of the required boundary term depends on the
choice of boundary conditions, and can be found by demanding the
existence of a classical extremum of the action.
\end{enumerate}

\section{Gauge Invariance and Chern-Simons Theory \label{CST}}

The problem of sewing path integrals becomes more interesting when one
starts with a theory with gauge invariance.  The simple archetype is
an abelian Chern-Simons theory on a three-manifold.  Let $A_\mu$ be an
abelian gauge potential (i.e., a connection on a $U(1)$ bundle over
$M$), and consider the action
\beq
I_M[A] = {k\over2\pi}
   \int_M \!d^3\sp x\,\epsilon^{\mu\nu\rho}A_\mu\partial_\nu A_\rho .
\label{b1}
\eeq
This action is invariant under gauge transformations
\beq
A_\mu\rightarrow A_\mu+\partial_\mu\Lambda ,
\label{b2}
\eeq
and leads to Euler-Lagrange equations
\beq
F_{\mu\nu} = \partial_\mu A_\nu - \partial_\nu A_\mu = 0 .
\label{b3}
\eeq
The space of classical solutions is thus the space of flat connections
modulo gauge transformations.  The corresponding quantum theory is
fairly simple, and it may be shown to have a finite-dimensional Hilbert
space (see, for example, \cite{Poly}).

We next split $M$ into two pieces along $\Sigma$ and repeat the
analysis of the preceding section.  The analogue of the surface term
in equation \rref{a5} is
\beq
\delta I_{M_1}[A] = \dots -{k\over2\pi}\int_\Sigma \!d^2\sp x\, n_\rho
   \epsilon^{\rho\mu\nu}A_\mu\delta A_\nu ,
\label{b4}
\eeq
and we must choose boundary conditions and a surface term $I_\Sigma$
to cancel this variation.  A standard approach is to choose a complex
structure on $\Sigma$ and to fix the component $A_z$, which is canonically
conjugate to $A_{\bar z}$.  The variation \rref{b4} can then be cancelled
by the variation of a boundary action
\beq
I_\Sigma[A] = {k\over2\pi}\int_\Sigma \!d^2\sp x\, A_zA_{\bar z} .
\label{b5}
\eeq

Observe now that the action
\beq
I_{M_1}'[A] = I_{M_1}[A] + I_\Sigma[A]
\label{b6}
\eeq
is no longer invariant under gauge transformations \rref{b2} unless
$\Lambda$ vanishes at the boundary.  We can make this noninvariance
explicit by decomposing $A_\mu$ as
\beq
A_\mu = \bar A_\mu + \partial_\mu\Lambda ,
\label{b7}
\eeq
where $\bar A_\mu$ is a gauge-fixed potential; then
\beq
I_{M_1}'[A] = I_{M_1}'[\bar A]
   + {k\over2\pi}\int_\Sigma \!d^2\sp x
   \left(\partial_z\Lambda\partial_{\bar z}\Lambda
   + 2 \bar A_z\partial_{\bar z}\Lambda \right) .
\label{b8}
\eeq
The ``would-be gauge transformation'' $\Lambda$ has thus become a
dynamical field on $\Sigma$, with an action that can be recognized
as a chiral Wess-Zumino-Witten action.  The partition function
correspondingly factorizes,
\beq
Z[M_1][\bar A_z]
   = Z_{\hbox{\scriptsize bulk}}[\bar A_z]
     Z_{\hbox{\scriptsize WZW}}[\bar A_z] .
\label{b9}
\eeq
This is a dramatic result: we have gone from a Chern-Simons quantum
theory with a finite-dimensional Hilbert space to a theory that
includes an infinite-dimensional Hilbert space describing boundary
degrees of freedom.

An analogous process occurs in the nonabelian case.  Let $A =
A_\mu^a T_a dx^\mu$ denote a connection one-form for a nonabelian
gauge group $G$ with generators $T_a$.  Then the Chern-Simons action
\beq
I_{M_1}'[A] = {k\over4\pi}\int_{M_1} \hbox{Tr}\left(
   A\wedge dA + {2\over3}A\wedge A\wedge A\right)
   + {k\over4\pi}\int_\Sigma \hbox{Tr}A_zA_{\bar z}
\label{b10}
\eeq
appropriate for fixing $A_z$ at $\Sigma$ again splits into two pieces;
under the decomposition
\beq
A = g^{-1}dg + g^{-1}\bar A g ,
\label{b11}
\eeq
the action becomes \cite{Ogura,CarWZW}
\beq
I_{M_1}'[\bar A,g]
   = I_{M_1}'[\bar A] + k I_{\hbox{\scriptsize WZW}}[g,\bar A_z] .
\label{b12}
\eeq
The term $I_{\hbox{\scriptsize WZW}}[g,\bar A_z]$ is now the action of
a nonabelian WZW model at the boundary $\Sigma$, and a decomposition
of the partition function of the form \rref{b9} again holds.

As in the case of a scalar field, the factor in $Z[M_1]$ coming from
the boundary term is necessary to ensure a sewing relation analogous
to \rref{a10}.  Witten has shown that when the WZW action is included,
Chern-Simons theory does indeed sew properly at boundaries \cite{WitWZW}.

Once again, none of this is new.  But we can add two more lessons to
the two at the end of the preceding section:
\begin{enumerate}
\setcounter{enumi}{2}
\item In a gauge theory, the presence of a boundary term can break the
gauge invariance, leading to new boundary degrees of freedom;
\item The new, dynamical ``would-be gauge'' degrees of freedom can
drastically alter the Hilbert space of the corresponding quantum
theory.
\end{enumerate}
This last observation is the key to the proposed explanation of black
hole entropy.

\section{(2+1)-Dimensional Gravity}

As our third example, let us examine quantum gravity in three spacetime
dimensions.  This model has the beautiful feature, first noticed by
Ach{\'u}carro and Townsend \cite{Achu} and later developed by Witten
\cite{Wita}, that it can be rewritten as a Chern-Simons theory.  In
particular, suppose that  a negative cosmological constant $\Lambda =
-1/\ell^2$ is present, as is appropriate for the black hole solution of
Ba{\~n}ados, Teitelboim, and Zanelli \cite{BTZ}.  Define the two
$SO(2,1)$ gauge fields
\beq
A^a = \omega^a + {1\over\ell}e^a , \quad
  \tilde A^a = \omega^a - {1\over\ell}e^a ,
\label{c1}
\eeq
where $e^a = e^a_\mu dx^\mu$ is a triad ($g_{\mu\nu} =
e^a_\mu e^b_\nu\eta_{ab}$ is the metric) and $\omega^a = {1\over2}
\epsilon^{abc}\omega_{\mu bc}dx^\mu$ is a spin connection.  The
first-order form of the standard Einstein-Hilberta action for general
relativity may then be written as
\beq
I_{\hbox{\scriptsize grav}}[e,\omega] = I[A] - I[\tilde A] ,
\label{c3}
\eeq
where $I[A]$ is the Chern-Simons action \rref{b10} with
\beq
k = {\ell\sqrt{2}\over 8G} .
\label{c4}
\eeq
(See reference \cite{Carbh} for my conventions.)  In this formulation,
the diffeomorphisms and local Lorentz transformations of ordinary
general relativity are transmuted into $SO(2,1)\!\times\!SO(2,1)$
gauge transformations, parametrized by the group elements $g$ as in
\rref{b11}.

Based on what we have seen of Chern-Simons theory, we might expect
a boundary Wess-Zumino-Witten action to again be induced on $\Sigma$.
In particular, if $\Sigma$ is the horizon of a black hole, the
boundary fields found above are candidates for the microscopic
degrees of freedom responsible for the black hole entropy.  The
boundary data $A_z$ and $\tilde A_z$ of the preceding section are
not quite suitable for this situation, but the appropriate data---which
specify that $\Sigma$ is an apparent horizon of a fixed circumference
$2\pi r_+$---may be shown to again give rise to an $SO(2,1)\!\times\!
SO(2,1)$ WZW action \cite{Carbh}.

Now, WZW models for noncompact groups such as $SO(2,1)$ are not yet
completely understood.  In the large $k$ (or small $\Lambda$) limit,
however, the $SO(2,1)\!\times\!SO(2,1)$ action may be approximated
by a system of six independent bosonic string oscillators.  Such a
system has an infinite number of states, but most of these are eliminated
by a remaining gauge symmetry---a remnant of the Wheeler-DeWitt
equation---that expresses invariance under shifts of the angular
coordinate $\phi$.  Given reasonable assumptions about the Hilbert
space upon which this symmetry acts, it is shown in reference
\cite{Carbh} that the number of boundary states is
\beq
n \sim \exp\left\{ {2\pi r_+\over4G}\right\} .
\label{c5}
\eeq
The logarithm of this expression gives the correct Bekenstein-Hawking
entropy \cite{BTZ},
\beq
S = {2\pi r_+\over4G} ,
\label{c6}
\eeq
for the (2+1)-dimensional black hole.

\section{(3+1)-Dimensional Black Holes}

Unfortunately, these results in 2+1 dimensions cannot be translated
directly to realistic (3+1)-dimensional gravity.  In particular, the
Chern-Simons formulation of (2+1)-dimensional gravity allows a clean
separation of ``physical'' and ``gauge'' degrees of freedom, and no
(3+1)-dimensional analogue is known.  There are, however, some
tantalizing hints of a similar mechanism:
\begin{enumerate}
\item The obvious (3+1)-dimensional analogues of the dynamical ``would-be
gauge degrees of freedom'' in Chern-Simons theory are the ``would-be
diffeomorphisms'' that do not preserve the location of the boundary.
Just as the boundary term \rref{b5} breaks the gauge invariance of
Chern-Simons theory, the boundary terms\footnote{The first of these
terms is standard, although I am not sure of the right generalization
to null surfaces; the second occurs at the bifurcation two-sphere of
a black hole \cite{Brown}.}
$$ \int_\Sigma \!d^3\sp x\,\sqrt{h} K \quad \hbox{and} \quad
\int_\Sigma \!d^3\sp x\,\sqrt{\sigma}\,n^\mu\nabla_{\!\mu} N$$
break the naive diffeomorphism invariance of general relativity.  As
Marolf has noted \cite{Marolf}, one must be careful about how the
boundary is specified, but it may be possible to obtain a WZW-like
induced boundary action.
\item A standard method for obtaining the physical degrees of freedom
in general relativity \cite{York} is to split fluctuations $h_{ij}$ of
the spatial metric as
$$ h_{ij} = h_{ij}^{TT} + (L\xi)_{ij} + {1\over3}g_{ij}h ,$$
$$ \hbox{with}\quad
   (L\xi)_{ij} = D_i\xi_j + D_j\xi_i - {2\over3} g_{ij}D_k\xi^k ,\quad
   D^j h_{ij}^{TT}=0 ,$$
where $D_i$ denotes the spatial covariant derivative.  One then argues
that once the gauge is fixed and the constraints are imposed, only the
transverse traceless components $h_{ij}^{TT}$ remain as physical
degrees of freedom.  In the presence of a boundary, however, this
decomposition must be altered: one must impose boundary conditions on
$\xi_i$ and $h_{ij}^{TT}$ in order to make the operator $L$ self-adjoint.
As far as I know, the decomposition of the metric on a spatial slice with
boundary has not been analyzed, but it is at least suggestive that the
splitting into physical and gauge degrees of freedom must change.
\item As Balachandran, Chandar, and Momen have emphasized \cite{Bal},
the smeared generators of diffeomorphisms in canonical general relativity,
$$ {\cal D}_\xi = -2\int \!d^3\sp x\,\xi_iD_j\pi^{ij} ,$$
are functionally differentiable---and thus represent genuine
invariances---only when the vector $\xi^i$ vanishes at the boundary.
When $\xi^i\ne0$ at the boundary, the would-be constraints instead give
rise to an algebra of edge observables, which should presumably have a
representation on a Hilbert space of edge states.  An analysis of similar
edge observables in Chern-Simons theory can, indeed, reproduce the WZW
states described in section \ref{CST}
\item Preliminary investigations of (3+1)-dimensional quantum gravity
in the loop variable approach have indicated the appearance of
boundary states \cite{Smolin,Baez}.  In particular, Smolin has found
that in Euclidean quantum gravity with certain self-dual boundary
conditions, the would-be diffeomorphisms of the boundary become
dynamical observables, which act on states that may be described in
terms of an induced topological field theory at the boundary.
\end{enumerate}

These results remain scattered, and do not yet give a full, coherent
picture of the origin of black hole entropy.  I believe, however,
that such a picture is emerging, and that we have cause for
optimism.

\vspace{1.5ex}
\begin{flushleft}
\large\bf Acknowledgements
\end{flushleft}

This work was supported in part by National Science Foundation grant
PHY-93-57203 and Department of Energy grant DE-FG03-91ER40674.

\end{document}